\documentclass[journal]{IEEEtran}

\usepackage{mathtools}
\usepackage{makecell}
\usepackage{threeparttable}
\usepackage{caption}
\usepackage{array}
\usepackage{booktabs}
\usepackage{multirow}
\usepackage[bookmarks=false]{hyperref}
\usepackage{pifont}

\usepackage{cite}
\usepackage{amsmath,amsfonts}
\usepackage{algorithmic}
\usepackage{graphicx}
\usepackage{textcomp}

\usepackage[ruled,vlined]{algorithm2e}

\usepackage{framed,multirow}

\usepackage{amssymb}
\usepackage{latexsym}

\usepackage{url}
\usepackage[dvipsnames]{xcolor}
\definecolor{mypink1}{rgb}{0.858, 0.188, 0.478}

\usepackage{soul} 

\begin{document}

\title{Structure-aware scale-adaptive networks for cancer segmentation in whole-slide images}

\author{Yibao Sun${}^{1}$, Giussepi Lopez${}^{1}$, Yaqi Wang${}^{2}$, Xingru Huang${}^{1\star}$, Huiyu Zhou${}^{3}$, Qianni Zhang${}^{1}$
\thanks{${}^\star$Corresponding author: Xingru Huang.}
\thanks{${}^{1}$Yibao Sun, Giussepi Lopez, Xingru Huang and Qianni Zhang are with the School of Electronic Engineering and Computer Science, Queen Mary University of London, Mile End Road, London, E1 4NS, United Kingdom (e-mail: yibao.sun; e.g.lopezmolina; xingru.huang; qianni.zhang@qmul.ac.uk).}
\thanks{${}^{2}$Yaqi Wang is with the College of Media Engineering, Communication University of Zhejiang, Hangzhou, China (e-mail: wangyaqi@cuz.edu.cn).}
\thanks{${}^{3}$Huiyu Zhou is with the School of Informatics, University of Leicester, University Road, Leicester, LE1 7RH, United Kingdom (e-mail: hz143@leicester.ac.uk).}
}

\maketitle

\begin{abstract}
Cancer segmentation in whole-slide images is a fundamental step for viable tumour burden estimation, which is of great value for cancer assessment. 
However, factors like vague boundaries or small regions dissociated from viable tumour areas make it a challenging task.
Considering the usefulness of multi-scale features in various vision-related tasks, we present a structure-aware scale-adaptive feature selection method for efficient and accurate cancer segmentation. Based on a segmentation network with a popular encoder-decoder architecture, a scale-adaptive module is proposed for selecting more robust features to represent the vague, non-rigid boundaries. Furthermore, a structural similarity metric is proposed for better tissue structure awareness to deal with small region segmentation. In addition, advanced designs including several attention mechanisms and the selective-kernel convolutions are applied to the baseline network for comparative study purposes. Extensive experimental results show that the proposed structure-aware scale-adaptive networks achieve outstanding performance on liver cancer segmentation when compared to top ten submitted results in the challenge of PAIP 2019. Further evaluation on colorectal cancer segmentation shows that the scale-adaptive module improves the baseline network or outperforms the other excellent designs of attention mechanisms when considering the tradeoff between efficiency and accuracy. The sourcecode is publicly available at: \textcolor{mypink1}{\url{https://github.com/sigma10010/histo_img_seg}}.
\end{abstract}


\section{Introduction}
According to the Global Cancer Statistics 2020 of cancer incidence and mortality produced by the International Agency for Research on Cancer, cancer is a leading cause of death worldwide, accounting for nearly 10 million deaths in 2020 \cite{sung2021global}.
Histopathology is considered a clinical gold standard for diagnosis of a variety of diseases including cancers. Surgical resection is the only treatment associated with long-term survival for patients with most types of cancer. Samples from the resected cancerous tissue are used to produce stained slides for further analysis. Measurements like tumour burden are obtained and used by radiologists to evaluate the prognosis after clinic surgery.

Image segmentation, in the context of computer vision, is defined as the process of assigning labels to all the pixels from an image; as a result, a mask covering the whole image is created.
In the medical field, several types of images are used to detect and diagnose cancer, e.g., computed tomographies, magnetic resonance images (MRIs), ultrasound scans or sonograms, digitized specimen slides; among others. 
Segmentation techniques have been successfully applied to several medical tasks like pneumothorax segmentation \cite{tolkachev2020deep}, brain tumour segmentation \cite{shen2017fully, akil2020fully}, skin lesions \cite{lei2020skin}, cardiac MRIs segmentation \cite{alba2018automatic}, etc. The promising outcomes have highlighted the importance of using computer-aided diagnosis (CAD) to assist the experts with the detection and quantitative analysis  of cancer.

Tumour area segmentation in stained histopathology tissue slides is a fundamental step for cancer grading and assessment when used along with CAD systems, especially for viable tumour burden quantification. Traditionally, experts use a semiquantitative grading system for analyzing the degree of tumour burden. However, manual delineation requires professional domain knowledge and is laborious and time-consuming due to the large variations in the shape of the tumour regions. Also, the standard of semiquantitative grading is based on the experience of doctors, which is unavoidably subjective and suffers from inter-observer variability, leading to uncertainty in examination of difficult cases. Therefore, automatic solutions for accurate, objective and reproducible tumour area segmentation in whole-slide images are desired for many applications, such as estimating the degree of tumour burden. 

Although automatic tumour area segmentation is highly demanded and many endeavours have been poured into, there still have several challenges to deal with. First of all, the boundary of tumour area is irregular and fuzzy due to the infiltration of cancerous cells. Existing methods try to learn more robust features to solve the issue \cite{schlemper2019attention}. In \cite{schmitz2021multi}, multi-scale features have been demonstrated to be potential in cancer segmentation. However, as it can be seen from their experimental results, the model performs unsatisfactorily and only secure the $7^{th}$ place in a challenge on hepatocellular carcinoma segmentation. Another challenge is that there always have some small regions similar to cancer in whole-slide images apart from the viable tumour areas. Previous approaches like U-Net \cite{ronneberger2015u} cannot segment these small regions properly, which will be demonstrated in the Experiments (Section \ref{sec:seg-exp}). In addition to the vague boundary and the small regions, there are some other difficulties such as image variations between samples even when they come from the same patient, image artifacts created when preparing and scanning the samples (wrinkles, dust, blur created by samples with different density, among others), colours variation when using products from different vendors, large dimensions of WSI (they are in the gigapixel range, consequently, they cannot be directly used as an input for a neuronal network), class imbalance (remarked difference of samples per class) and distribution imbalance (large difference in the areas covered by each class), etc.

To address these challenges for the purpose of efficient and accurate cancer segmentation, we present structure-aware scale-adaptive feature selection networks, by leveraging state-of-the-art techniques of deep learning with tailored designs including attention mechanisms, residual learning \cite{he2016deep}, structural similarity and multi-scale feature fusion, etc. Contributions includes: 1) We design a scale-adaptive module for dynamic feature selection, which is easily integrated into a segmentation network and can ensure the network to learn more robust features around tumour for the vague boundary problem. Instead of averagely fusing features from different scales, we propose to learn weights for scale-adaptive feature selection; 2) A structural similarity metric is proposed to regulate the network training procedure for better tissue structure awareness, which is helpful for solving the issue of small regions. We empirically validate the impact of the structural similarity loss on regulating networks training and its improvement on tumour area segmentation; 3) We comprehensively study and compare the proposed structure-aware scale-adaptive networks with several advanced designs of attention mechanisms on tumour area segmentation. All the sourcecode has been made publicly available online.

\section{Related work}
As convolutional neural networks (CNN) become the state-of-the-art in learning visual representation, it is natural to apply CNNs to solve the image segmentation problem for the requirement of coarse-to-fine inference. Inspired by the process of using CNNs for image classification in which a category is assigned to each input image, early segmentation approaches directly adopt the classification networks to predict a category for each pixel through training a patch-wise classifier with much smaller input size when compared to the whole input image \cite{farabet2012learning, gatta2014unrolling}. This kind of patch-based methods always work with a sliding window to obtain segmentation map. Typically, a patch is fed into the trained classifier to predict a label for the central pixel of the patch. The patch-based methods for semantic segmentation have inherent drawbacks like low computational efficiency and blocky prediction result \cite{grangier2009deep}. 

To address the drawbacks of patch-based methods, fully convolutional networks (FCN) have been proposed since 2015 \cite{long2015fully}, which has driven recent advances in applying CNN for image segmentation. Many works including U-Net \cite{ronneberger2015u}, SegNet \cite{badrinarayanan2017segnet}, DeconvNet \cite{noh2015learning} and Deeplab \cite{chen2017deeplab} improve FCN with a better encoder-decoder network architecture.
In the field of medical imaging, an increasing number of U-Net variants have been successfully applied to various scenarios and image modalities \cite{zhou2018unet++,li2019cascade,li2020attention}, including digitised histopathological images \cite{li2020net,feng2020deep,shah2021colorectal}.
The main advantage of these FCN based methods for pixel labelling when compared to the patch-based ones is that the networks can be trained in an end-to-end, pixel-to-pixel way.

Due to the impressive success of CNN for natural image classification and segmentation, there has been an increasing interest in applying this new technology to histology image analysis \cite{janowczyk2016deep,litjens2017survey}. For instance, authors in \cite{araujo2017classification} propose an convolutional neural network to extract information from multiple scales to classify breast cancer histology images and their approach achieves a high sensitivity of 95.6\% for carcinoma cases. Deep convolutional neural networks are also used for classifying histology images of many other kinds of cancer \cite{sharma2017deep}. Besides histopathology image classification, there are also strong attempts for histology image segmentation. Ciresan et al. propose a network in a sliding-window setup to predict a class label for each pixel, which won the electron microscopy segmentation challenge at ISBI 2012 \cite{ciresan2012deep}. Most early methods for cancer segmentation are empowered by patch-based classifier with a sliding-window post processing. As FCN becomes the state-of-the-art in image segmentation, many work adopt the encoder-decoder architecture for end-to-end cancer segmentation \cite{jahromi2020deep,schmitz2021multi,wang2021hybrid}. Among them, an ensemble model achieves the best published result on hepatocellular carcinoma segmentation \cite{wang2021hybrid}. Following these researches, we also adopt the encoder-decoder architecture in our design. A comparative analysis along with these previous methods is given in Section \ref{sec:paip2019}.

Inspired by the important role of attention mechanism in human visual perception, many early researches have made success in applying visual attention on scene understanding \cite{itti1998model,rensink2000dynamic}. Due to the success of deep learning \cite{lecun2015deep}, especially the breakthroughs of CNN in learning powerful visual representation \cite{krizhevsky2012imagenet,szegedy2015going,he2016deep}, an increasing amount of works try to leverage attention mechanism for further improvement on CNN \cite{fu2017look,hu2018squeeze,fu2019dual,mi2020hierarchical,guo2021ssan}. Among them, several attention mechanisms including the squeeze-and-excitation (SE) block \cite{hu2018squeeze}, the bottleneck attention module (BAM) \cite{park2018bam} and the convolutional block attention module (CBAM) \cite{woo2018cbam} gain great success in more robust feature learning for a couple of visual tasks including image segmentation. Some works like the attention gated networks attempt to leverage salient regions for medical image segmentation \cite{oktay2018attention,schlemper2019attention} and achieve the state-of-the-art results. We are going to compare the proposed network with these advanced networks based on attention mechanisms.

\section{Methodology}
\label{sec:method}
Inspired by the recent advances in structural similarity and hierarchical feature fusion for better visual representation learning for semantic segmentation, we propose a structure-aware scale-adaptive feature selection network for cancer segmentation in whole-slide images. In this section, we describe in details the proposed networks and loss functions.

\subsection{Overview of the network architecture}
\label{sec:archi}
\begin{figure*}
\centering
\includegraphics[width=0.92\textwidth]{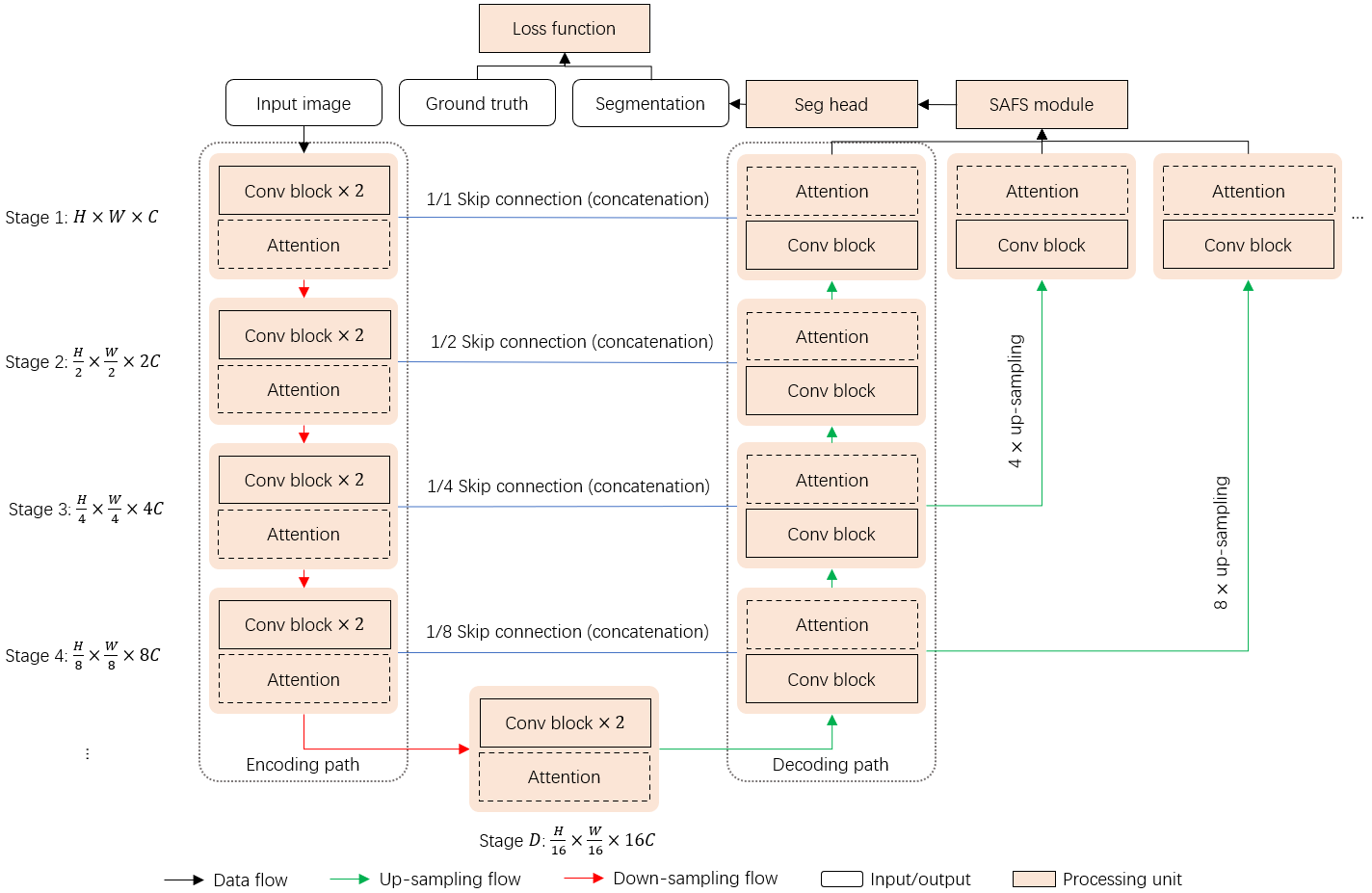}
\caption{\label{fig:ssunet}The architecture of the proposed structure-aware scale-adaptive networks. For concise purpose, down-sampling units between each two stages in the encoding path and up-sampling units in the decoding path are omitted. Processing units in dashed boxes denote that attention blocks are optional.}
\end{figure*}
\begin{figure}
\centering
\includegraphics[width=0.48\textwidth]{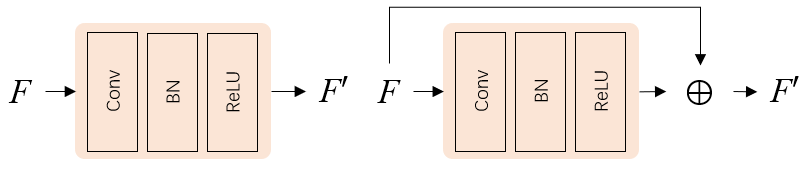}
\caption{\label{fig:conv} Building blocks. Left: a basic convolutional block. Right: a basic convolutional block with shortcut. F: feature map, Conv: convolution, BN: batch normalization \cite{ioffe2015batch}, ReLU: rectified linear unit \cite{nair2010rectified}.}
\end{figure}

Beyond the patch-based methods, U-Net and its variants form a major stream in medical image segmentation. According to the summary of the challenge on liver cancer segmentation \cite{kim2021paip}, most of the submitted top methods apply U-Net variants with a similar encoder-decoder architecture for tumour area segmentation.
For the purpose of comparative analysis, we also adopt an encoder-decoder architecture for the proposed networks,
which is mainly composed of an encoding path and a decoding path, as illustrated in Fig. \ref{fig:ssunet}. In the encoding path, basic convolutional blocks (Fig. \ref{fig:conv}) and down-sampling units are stacked for feature map extraction. The feature map output via the encoding path is then gradually decoded in the decoding path which consists of convolutional blocks and up-sampling units. The decoded feature map at the final stage, of the same size as the input image, is then fed to a segmentation head - a convolution with $1 \times 1$ kernel - for final segmentation prediction. Instead of using a single feature map for prediction, multi-scale feature maps decoded at different stages are simultaneously fed to a scale-adaptive feature selection (SAFS) module before being sent to the segmentation head. More details of the SAFS module are given in Section \ref{sec:ssm}.

\subsection{Network variants}
For the purpose of comparative studies, we instantiate several networks with different settings in terms of skip/shortcut connections, convolutional blocks, and attention blocks, etc. The first one is a baseline U-Net, which is built of basic convolutional blocks without any attention block applied, and only a single feature map is fed to the segmentation head for prediction. The second one is based on U-Net but built of shortcut convolutional blocks, as shown in Fig. \ref{fig:conv}, referred to as U-Net+SC to distinguish from the baseline U-Net. 

Several state-of-the-art attention mechanisms including the attention gate (AG) \cite{oktay2018attention,schlemper2019attention}, the squeeze-and-excitation (SE) block  \cite{hu2018squeeze}, the bottleneck attention module (BAM) \cite{park2018bam} and the convolutional block attention module (CBAM) \cite{woo2018cbam} are integrated into the baseline U-Net, the resulting networks are named as U-Net+AG, U-Net+SE, U-Net+BAM and U-Net+CBAM respectively.
Models trained with AG implicitly learn to suppress irrelevant regions in an input image while highlighting useful salient features. It is added to each layer in the decoding path of the baseline U-Net.
SE applies only channel attention, while both BAM and CBAM combine channel attention and spatial attention in a sequential way,
yet there are some slight differences on how to aggregate information across spatial dimension between them.
All of SE, BAM and CBAM are applied to every intermediate feature map of each stage in both the encoding path and the decoding path.

Since dependencies among tissue types normally exist in a whole-slide range, being able to capture long-range spatial dependencies is critical for cancer segmentation. This is, however, not possible for CNN due to its property of local connectivity. A novel selective-kernel (SK) convolution has been proposed for larger and adaptive receptive fields to improve CNN \cite{li2019selective}.
To exploit the effectiveness of the adaptive receptive fields on histopathological image segmentation, the basic convolutional blocks in the encoding path of the baseline U-Net are substituted as the selective-kernel convolutions to form a new network, which is named as U-Net+SK.

\subsection{The scale-adaptive feature selection module}
\label{sec:ssm}
Multi-scale features have been demonstrated very useful for many vision related tasks such as object detection \cite{lin2017feature}, MRI reconstruction \cite{nguyen2019frequency} and histopathological image segmentation \cite{schmitz2021multi}. To explore its effectiveness in tumour area segmentation, we design a network called U-Net+MS, in which multiple feature maps $\{F_1, F_2,...,F_n\}$ decoded at different stages of different scales are sent to the segmentation head in a parallel way for final prediction. A simple fusion is performed to acquire an averaged feature map $\overline{F}$, i.e., $\overline{F} = \frac{1}{n}\sum_{i=1}^n F_i $, before prediction. The averaged feature map can take the advantages of multi-scale features. Meanwhile, it could degrade the overall performance when features from some specific scales have poor performance. Inspired by the work of selective kernel for adaptive receptive fields of CNN \cite{li2019selective}, we propose a scale-adaptive feature selection (SAFS) module as an alternative feature fusion method, which is integrated into the baseline U-Net. The resulting network is referred to as U-Net+SAFS in the following descriptions.

\begin{figure*}
\centering
\includegraphics[width=0.8\textwidth]{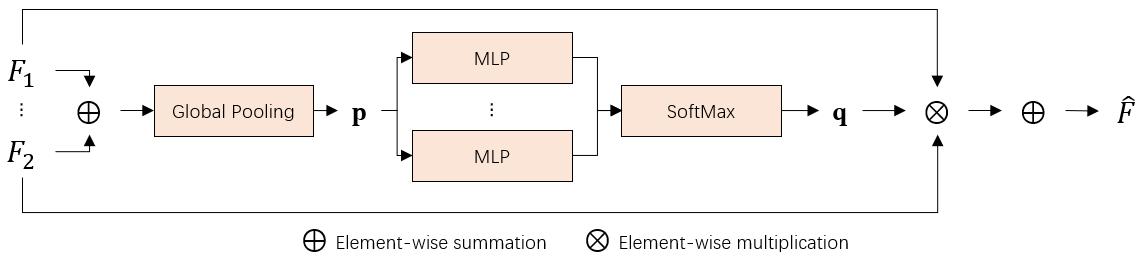}
\caption{\label{fig:ss} The scale-adaptive feature selection module. MLP: multi-layer perceptron with a single hidden layer.}
\end{figure*}

A two-branch case of the SAFS module is shown in Fig. \ref{fig:ss}. Given feature maps $F_1$ and $F_2$ from different scales, which are fused together to obtain a united feature map $F$, i.e., $F=F_1 \oplus F_2$, where $\oplus$ is an element-wise summation. The united feature map $F \in \mathbb{R}^{C \times H \times W}$ is then fed to a global pooling unit to generate channel-wise statistics $\mathbf{p}= \{p_i | i = 1,2,...,C\}$, where $p_i=\frac{1}{H \times W}\sum_{r=1}^{H}\sum_{c=1}^{W}F(r,c)$, which aggregates information across spatial dimensions. The summation fusion procedure before the global pooling unit ensures the information is aggregated from all the branches of different scales. Multiple paralleled MLPs and a SoftMax activation are used to map the statistics $\mathbf{p}$ to a concatenated channel attention weights $\mathbf{q}$ which can be expressed as $\mathbf{q}= \mathbf{a} \cup \mathbf{b}$, where $\mathbf{a}$ and $\mathbf{b}$ are channel attention weights for feature maps from different scales respectively, i.e., $\mathbf{a} = \{a_i|i=1,2,...,C\}$, $\mathbf{b}=\{b_i|i=1,2,...,C\}$, which subjects to $a_i+b_i=1$. Finally, a scale-adaptive feature map $\hat{F}$ is obtained via a dynamic selection procedure, which can be expressed as $\hat{F}=(\mathbf{a} \otimes F_1) \oplus (\mathbf{b} \otimes F_2)$,  where $\otimes$ is an element-wise multiplication. For the sake of clarity, these steps are summarised in the Algorithm \ref{alg:safs}. As it can be seen, the scale-adaptive feature is actually a weighted average of features at different scales. Unlike an averagely fused feature, the scale-adaptive feature takes advantages of features at different scales without performance degradation, since the weights are learned in a self-adaptive way.

\begin{algorithm}
\caption{\label{alg:safs} The scale-adaptive feature selection}
\begin{algorithmic}[1]
\REQUIRE Feature maps $\left \{  F_{i}  | i=1,2,...,n \right \}$, s.t. $F_i \in \mathbb{R}^{C \times H \times W}$
\ENSURE A scale-adaptive feature map $\hat{F}$
\STATE Concatenate feature map: $\tilde{F} \leftarrow Cat(F_1,F_2,...F_n)$, s.t. $\tilde{F} \in \mathbb{R}^{n \times C \times H \times W}$
\STATE Unite feature map: $F \leftarrow \sum_{i=1}^n F_i$, s.t. $F \in \mathbb{R}^{C \times H \times W}$
\STATE Spatial information aggregation: $\mathbf{p} \leftarrow \frac{1}{H \times W}\sum_{r=1}^{H}\sum_{c=1}^{W}F(r,c)$, s.t. $\mathbf{p} \in \mathbb{R}^C$
\STATE Reduction rate: $r \leftarrow 8$
\STATE Define a fully connected layer: $FC(in\_dims: C, out\_dims: \frac{C}{r})$
\STATE Dimension reduction: $\mathbf{z} \leftarrow FC(\mathbf{p})$, s.t.  $\mathbf{z} \in \mathbb{R}^{\frac{C}{r}}$
\STATE $\mathbf{z'} \leftarrow \{\}$
\WHILE{$n \neq 0$}
\STATE Define a fully connected layer: $FC(in\_dims: \frac{C}{r}, out\_dims: C)$
\STATE $\mathbf{z'} \leftarrow \mathbf{z'} \cup FC(\mathbf{z})$
\STATE $n \leftarrow n - 1$
\ENDWHILE
\STATE $\mathbf{q} \leftarrow SoftMax(\mathbf{z'})$, s.t. $\mathbf{q} \in \mathbb{R}^{n \times C}$
\STATE Expand $\mathbf{q}$, s.t. $\mathbf{q} \in \mathbb{R}^{n \times C \times H \times W}$
\STATE $\hat{F} \leftarrow Sum(\mathbf{q} \otimes \tilde{F}, dim=0)$, s.t. $\hat{F} \in \mathbb{R}^{C \times H \times W}$, $\otimes$ denotes element-wise multiplication.
\end{algorithmic}
\end{algorithm}

\subsection{Loss functions}
As depicted in the network architecture (Fig. \ref{fig:ssunet}), given an image as the input, a neural network predicts a segmentation map of the same size as the input image. Widely used cross-entropy loss is applied to train networks with different configurations. Given a prediction map $M^{(p)} \in \mathbb{R}^{H \times W}$ and the corresponding ground-truth map $M^{(g)}\in \mathbb{R}^{H \times W}$ of size of $H \times W$ pixels, the pixel-wise cross-entropy loss can be expressed as:

\begin{equation}
\begin{aligned}
\mathcal{L}_{ce}(M^{(p)}, M^{(g)})=-\frac{1}{H \times W}\sum_{r=1}^{H}\sum_{c=1}^{W}[ M^{(g)}(r,c) log( M^{(p)}(r,c))&\\
+(1-M^{(g)}(r,c)) log(1-M^{(p)}(r,c))]&
\end{aligned}
\end{equation}
where $M^{(g)}(r,c) \in \{0,1\}$ is the ground truth label at pixel $(r,c)$, and $M^{(p)}(r,c) \in [0,1]$ is the predicted probability of being target instance.

Pixels located around tumour border play a key role in tumour area segmentation. To focus on those key pixels, structural similarity is applied to capture local differences between a prediction map and the corresponding ground-truth map.
A structural similarity (SSIM) index has been proposed for perceived image quality assessment, which compares luminance, contrast and structure between two images \cite{zhou2004image}.
Let $\mathbf{x}=\{\mathbf{x}_i | i=1,2,...,K^2 \}$ denote a patch of $K \times K$ pixels extracted from a predicted segmentation map, i.e., $\mathbf{x} \subseteq M^{(p)}$, and $\mathbf{y}=\{\mathbf{y}_i | i=1,2,...,K^2 \}$ denote the aligned patch of the corresponding ground-truth map $M^{(g)}$,
the SSIM index of $\mathbf{x}$ and $\mathbf{y}$ can then be calculated as:
\begin{equation}
\begin{aligned}
SSIM(\mathbf{x}, \mathbf{y}) = \frac{(2\mu_x\mu_y+C_1)(2\sigma_{xy}+C_2 )}{(\mu_x^2+\mu_y^2+C_1)(\sigma_x^2+\sigma_y^2+C_2)}
\end{aligned}
\end{equation}
where $\mu_x$ and $\mu_y$ are the mean intensity of $\mathbf{x}$ and $\mathbf{y}$ respectively, $\sigma_x$ and $\sigma_y$ are the standard deviation of $\mathbf{x}$ and $\mathbf{y}$ respectively, $\sigma_{xy}$ is the covariance between $\mathbf{x}$ and $\mathbf{y}$, and $C_1$, $C_2$ are small constants to avoid instability when $\mu_x^2+\mu_y^2$ or $\sigma_x^2+\sigma_y^2$ is very close to zero, which are set as $0.01^2$ and $0.03^2$ respectively in our experiments.

With the above definitions, the SSIM loss between a prediction map $M^{(p)} \in \mathbb{R}^{H \times W}$ and the corresponding ground-truth map $M^{(g)}\in \mathbb{R}^{H \times W}$ is calculated as:

\begin{equation}
\begin{aligned}
\mathcal{L}_{ssim}&(M^{(p)}, M^{(g)})=\\
&\frac{1}{(H-K)(W-K)}\sum_{\forall \mathbf{x}\subseteq M^{(p)}, \forall \mathbf{y}\subseteq M^{(g)}}(1-SSIM(\mathbf{x}, \mathbf{y}))
\end{aligned}
\end{equation}
where $(H-K)(W-K)$ is the total number of patch pair $(\mathbf{x}, \mathbf{y})$ to calculate $SSIM$. In our experiments, we choose $K=11$ and apply a sliding window (kernel) with a stride of 1 to extract patches. Moreover, we apply multi-scale structural similarity \cite{wang2003multiscale}, which is more flexible than the single-scale one.
It can be observed that the SSIM loss is calculated on small patches extracted from prediction map and ground truth. Losses between patches located within areas of tumour or other tissue types are small while those of patches around tumour border are large. To minimize the SSIM loss forces the networks to focus on pixels around tumour border, so as to improve the segmentation results. This is demonstrated in the ablation study (Section \ref{sec:ablation}).

Besides the SSIM loss, a couple of metric based losses have been proposed for image classification and segmentation, such as the IoU loss \cite{rahman2016optimizing,mattyus2017deeproadmapper} and the Dice loss \cite{sudre2017generalised}. In order to perform comparative study, we also test the IoU loss, which in the same context as the cross-entropy loss can be expressed as:

\begin{equation}
\begin{aligned}
&\mathcal{L}_{iou}(M^{(p)}, M^{(g)})=\\
&1-\frac{\sum_{r=1}^{H}\sum_{c=1}^{W}M^{(g)}(r,c)M^{(p)}(r,c)}{\sum_{r=1}^{H}\sum_{c=1}^{W}[M^{(g)}(r,c)+M^{(p)}(r,c)-M^{(g)}(r,c)M^{(p)}(r,c)]}
\end{aligned}
\end{equation}
Different to the SSIM loss, the IoU loss is a global measure between the segmentation map and the ground truth.

\section{Experiments}
\label{sec:seg-exp}
\subsection{Datasets}
\begin{table*}[]
\centering
\begin{threeparttable}
\caption{A summary of the datasets and the derived image patches.}
\label{tab:paip}
\begin{tabular}{l|l|c|c|c|c}
\Xhline{3\arrayrulewidth}
\multicolumn{2}{l|}{Dataset} & \multicolumn{3}{c|}{PAIP 2019} & PAIP 2020 \\ \Xhline{2\arrayrulewidth}
\multirow{2}{*}{\rotatebox{90}{WSI}} & Type & \multicolumn{3}{c|}{Liver cancer} & Colorectal cancer \\ \cline{2-6} 
 & Amount & \multicolumn{3}{c|}{50} & 47 \\ \hline
\multirow{3}{*}{\rotatebox{90}{Patch}} & Size & 256 $\times$ 256 & 400 $\times$ 400 & 512 $\times$ 512 & 384 $\times$ 384 \\ \cline{2-6} 
 & Level & 2 & 2 & 2 & 3 \\ \cline{2-6} 
 & Amount & 15,637 & 6,161 & 5,850 & 4,772 \\ \Xhline{3\arrayrulewidth}
\end{tabular}
\end{threeparttable}
\end{table*}
Two published datasets (PAIP 2019 and PAIP 2020) for liver cancer and colorectal cancer assessment are used for evaluation. Table \ref{tab:paip} gives a summary for whole-slide images (WSIs) and patches of two datasets respectively.
The PAIP 2019 \footnote{\href{https://paip2019.grand-challenge.org/Dataset/}{https://paip2019.grand-challenge.org/Dataset/}} dataset is released for liver cancer segmentation and viable tumour burden estimation, which contains 50 slides of liver cancer in total. All the WSIs are scanned by Aperio AT2 at $20 \times$ magnification (level 0). A ground-truth binary mask to denote viable tumour area or not is provided for each slide. Due to high resolution of whole-slide images and memory limitation, each whole-slide image is cropped into small patches of different sizes, e.g., $400 \times 400$ pixels, at the lowest magnification factor (level 2), with an overlap of 200 pixels for training purpose.
The PAIP 2020 \footnote{\href{https://paip2020.grand-challenge.org/Dataset/}{https://paip2020.grand-challenge.org/Dataset/}} dataset is released for the automated classification of molecular subtypes in colorectal cancer. There are 47 slides of colorectal cancer in total with annotated binary masks. All the WSIs are scanned by Aperio AT2 at $40 \times$ magnification. In total, 4,772 small patches each of $384 \times 384$ pixels are extracted from whole-slide images at the lowest magnification factor of $5 \times$ for training.
To test the proposed methods, whole-slide images are randomly split into 5 groups for 5-fold cross validation before they are cropped into small patches. Same split over WSIs is performed on the cropped patches to avoid correlation.

\subsection{Evaluation metrics}
Widely used metrics including the Jaccard similarity and the Dice coefficient are used to validate the effectiveness of the proposed method for cancer segmentation. Given a segmentation map ($M^{(s)}$) and its corresponding ground-truth map ($M^{(g)}$), the pixel-wise Jaccard similarity ($JS$)
and Dice coefficient ($DC$)
can be calculated as follows:
\begin{equation}
\begin{aligned}
JS(M^{(s)},M^{(g)})=\frac{\left | M^{(s)} \cap M^{(g)} \right |}{\left | M^{(s)} \cup  M^{(g)} \right |},
\end{aligned}
\end{equation}
\begin{equation}
\begin{aligned}
DC(M^{(s)},M^{(g)})=\frac{2\times \left  | M^{(s)} \cap M^{(g)} \right |}{\left | M^{(s)} \right | + \left | M^{(g)} \right |},
\end{aligned}
\end{equation}
where $\left | \cdot  \right |$ denotes the number of pixels labeled as target instance.
Besides the Jaccard similarity and the Dice coefficient, we also report specificity ($SP=\frac{TN}{TN+FP}$), sensitivity/recall ($RC=\frac{TP}{TP+FN}$) and precision ($PC=\frac{TP}{TP+FP}$)
for each test to provide additional evaluation, where $TP$, $FP$, $TN$, $FN$ stands for pixel-wise true positives, false positives, true negatives and false negatives respectively.

Normally, a valid evaluation of a model needs to base on a sequence of images. Therefore, given segmentation results and the corresponding ground-truth maps for the image sequence, we calculate the accumulated value of $TP$, $FP$, $TN$, $FN$ based on a pixel-wise confusion matrix. In cases of cancer segmentation in histopathological images, we usually have a dataset of image tiles and WSIs. For image tile segmentation, accumulated value of each metric on the testing dataset of tiles is considered as model performance. For WSI segmentation, a clipped Jaccard similarity is considered to penalize inaccurate result, which can be calculated as:
\begin{equation}
\begin{aligned}
\hat{JS_i}=\begin{cases}
 \frac{\left | M_i^{(s)} \cap M_i^{(g)} \right |}{\left | M_i^{(s)} \cup  M_i^{(g)} \right |},& \text{ if } \frac{\left | M_i^{(s)} \cap M_i^{(g)} \right |}{\left | M_i^{(s)} \cup  M_i^{(g)} \right |}\geq 0.65 \\ 
 0. & \text{ otherwise}
\end{cases}
\end{aligned}
\end{equation}
where $M_i^{(s)}$ denotes the segmentation map for the $i^{th}$ WSI and $M_i^{(g)}$ the corresponding ground-truth map. An average of $\Hat{JS}$ is considered as a final score on a testing dataset of WSIs:
\begin{equation}
\begin{aligned}
S_{wsi} = \frac{1}{N}\sum_{i=1}^N \hat{JS_i}
\end{aligned}
\end{equation}
where $N$ is the number of testing WSIs.

\subsection{Training and inference}
All the models are trained using the Adam \cite{kingma2014adam} optimizer with a basic learning rate of 1e-4 which is adjusted at last 10 training epochs with a decayed rate of 0.1. The coefficients for computing running averages of gradient and its square are set as 0.5 and 0.999 respectively. According to the losses convergence observation, we stop training after 100 training epochs. The training procedure takes around 5 hours for each model on a GPU of NVIDIA GeForce GTX Titan X, which depends on the range of batch size from 2 to 4.

To ensure the segmentation models are more robust against variations of histology images, data augmentation is performed during training by randomly transforming the training images in ways of colour jitter, rotation, horizontal flip and vertical flip with a certain probability of 0.5. For colour jitter, the image colour are randomly adjusted in their brightness, contrast, saturation and hue. For rotation, the images are rotated by a degree randomly selected from \{0,90,180,270\}. In the inference phase, evaluation on image patches and WSIs are performed respectively.

\subsection{Ablation study}
\label{sec:ablation}
To acquire an optimal setting for building networks for image segmentation, we extensively validate the effects of model scale, number of skip connection, amount of prediction head and different loss functions on liver cancer segmentation in terms of Dice coefficient ($DC$), Jaccard similarity ($JS$), score on whole-slide image ($S_{wsi}$), etc. 

Following \cite{tan2019efficientnet}, resolution, depth and width of networks are considered as three factors for measuring model scale. We define the network resolution as the size of input image ($H \times W$), depth as the last down-sampling stage of feature map ($D$) and width as the initial channel amount ($C$), as shown in the network architecture (Fig. \ref{fig:ssunet}). The performance of the baseline U-Net tested on the PAIP 2019 dataset against different scales of model are listed in Table \ref{tab:model-scale}. As can it be observed, the model performs better at the middle-level resolution of $400 \times 400$ on both image patches and WSIs when compared to the one of resolution of $256 \times 256$ or $512 \times 512$. For the depth factor, the performance increase as the network gets deeper, which is consistent with the common understanding of deep learning. Considering the trade-off between performance and model size, we choose $D=5$ for the rest of experiments. For the width factor, we test networks with different initial channels (16, 32 and 48). An observation is that the network with a width of 32 performs better on image patches while the one with a greater width of 48 gets a better result on WSIs. $C=32$ is chosen for the rest of the experiments when an additional condition, memory limitation, is considered.

\begin{table*}[]
\centering
\begin{threeparttable}
\caption{Effect of model scale (resolution, depth and width) on the performance of the U-Net.}
\label{tab:model-scale}
\begin{tabular}{l|c|c|c|c|c|c|c}
\Xhline{3\arrayrulewidth}
 & Model scale & $SP$ & $PC$ & $RC$ & $DC$ & $JS$ & $S_{wsi}$ \\ \Xhline{2\arrayrulewidth}
\multirow{3}{*}{\begin{tabular}[c]{@{}l@{}}\textbf{Resolution}\\ {[}Depth 5\\ Width 32{]}\end{tabular}} & $256 \times 256$ & 85.87 $\pm$ 2.99 & 84.90 $\pm$ 2.94 & \textbf{88.58 $\pm$ 7.27} & 86.59 $\pm$ 4.57 & 76.63 $\pm$ 6.83 & 31.08 $\pm$ 8.36 \\ 
 & \textbf{400 $\times$ 400} & \textbf{95.94 $\pm$ 1.08} & \textbf{91.22 $\pm$ 1.90} & 87.61 $\pm$ 7.02 & \textbf{89.24 $\pm$ 4.19} & \textbf{80.82 $\pm$ 6.68} & \textbf{73.78 $\pm$ 5.15} \\ 
 & $512 \times 512$ & 89.19 $\pm$ 3.20 & 88.12 $\pm$ 2.38 & 87.36 $\pm$ 6.85 & 87.64 $\pm$ 4.25 & 78.24 $\pm$ 6.45 & 34.34 $\pm$ 9.99 \\ \hline
\multirow{3}{*}{\begin{tabular}[c]{@{}l@{}}\textbf{Depth}\\ {[}Width 32\\ 400 $\times$ 400{]}\end{tabular}} & 3 & 91.28 $\pm$ 2.42 & 81.59 $\pm$ 3.50 & 79.62 $\pm$ 7.56 & 80.43 $\pm$ 4.81 & 67.54 $\pm$ 6.86 & 43.14 $\pm$ 9.63 \\ 
 & 4 & 94.36 $\pm$ 1.71 & 87.93 $\pm$ 3.44 & 85.73 $\pm$ 5.82 & 86.79 $\pm$ 4.52 & 76.94 $\pm$ 7.17 & 66.01 $\pm$ 7.39 \\ 
 & \textbf{5} & \textbf{95.94 $\pm$ 1.08} & \textbf{91.22 $\pm$ 1.90} & \textbf{87.61 $\pm$ 7.02} & \textbf{89.24 $\pm$ 4.19} & \textbf{80.82 $\pm$ 6.68} & \textbf{73.78 $\pm$ 5.15}   \\ \hline
\multirow{3}{*}{\begin{tabular}[c]{@{}l@{}}\textbf{Width}\\ {[}Depth 5\\ 400 $\times$ 400{]}\end{tabular}} & 16 & 95.09 $\pm$ 0.86 & 89.42 $\pm$ 2.14 & 87.29 $\pm$ 5.91 & 88.29 $\pm$ 3.99 & 79.26 $\pm$ 6.30 & 69.37 $\pm$ 6.15 \\ 
 & \textbf{32} & \textbf{95.94 $\pm$ 1.08} & \textbf{91.22 $\pm$ 1.90} & 87.61 $\pm$ 7.02 & \textbf{89.24 $\pm$ 4.19} & \textbf{80.82 $\pm$ 6.68} & 73.78 $\pm$ 5.15   \\ 
 & 48 & 95.60 $\pm$ 0.93 & 90.59 $\pm$ 1.38 & \textbf{87.77 $\pm$ 7.81} & 88.98 $\pm$ 4.51 & 80.44 $\pm$ 7.02 & \textbf{77.17 $\pm$ 7.83} \\ \Xhline{3\arrayrulewidth}
\end{tabular}
\end{threeparttable}
\end{table*}

To explore the skip-connection design for neural network construction, we test the baseline U-Net with different types of skip connection. Experimental results are listed in Table \ref{tab:skip}. All the networks with skip connection outperform the one without skip connection on WSIs (see the $S_{wsi}$ column), demonstrating that the skip-connection design is effective, although the improvement on image patches are not significant. In addition, networks with heavier skip connections are stronger than the ones with light skip connections. For example, when all feature maps at different stages (1/8+1/4+1/2+1/1) in the encoding path are connected and fused to the corresponding feature maps in the decoding path, the network achieves a better performance while compared against others.

\begin{table*}[]
\centering
\begin{threeparttable}
\caption{Effect of different types of skip connection on the performance of the U-Net.}
\label{tab:skip}
\begin{tabular}{l|c|c|c|c|c|c}
\Xhline{3\arrayrulewidth}
\# skip connection & $SP$ & $PC$ & $RC$ & $DC$ & $JS$ & $S_{wsi}$ \\ \Xhline{2\arrayrulewidth}
0 (without) & 94.73 $\pm$ 0.91 & 88.87 $\pm$ 2.38 & \textbf{87.83 $\pm$ 6.39} & 88.28 $\pm$ 4.34 & 79.29 $\pm$ 6.86 & 69.22 $\pm$ 8.86 \\
1/8 & 94.77 $\pm$ 2.06 & 89.13 $\pm$ 3.12 & 87.26 $\pm$ 7.21 & 88.04 $\pm$ 4.54 & 78.92 $\pm$ 7.11 & 70.70 $\pm$ 9.78\\
1/8+1/4  & 95.01 $\pm$ 1.13 & 89.51 $\pm$ 1.53 & 86.66 $\pm$ 7.74 & 87.90 $\pm$ 4.40 & 78.68 $\pm$ 6.93 & \textbf{75.92 $\pm$ 3.68}\\
1/8+1/4+1/2 & 94.93 $\pm$ 1.45 & 89.29 $\pm$ 2.46 & 87.26 $\pm$ 5.76 & 88.20 $\pm$ 3.86 & 79.10 $\pm$ 6.12  & 72.96 $\pm$ 6.73\\ 
1/8+1/4+1/2+1/1 & \textbf{95.94 $\pm$ 1.08} & \textbf{91.22 $\pm$ 1.90} & 87.61 $\pm$ 7.02 & \textbf{89.24 $\pm$ 4.19} & 8\textbf{0.82 $\pm$ 6.68} & 73.78 $\pm$ 5.15\\
\Xhline{3\arrayrulewidth}
\end{tabular}
\end{threeparttable}
\end{table*}

To observe the effect of different fusion schemes for multi-scale features in prediction, feature maps at different decoding stages are fused for final prediction. Two kinds of fusion methods are tested, i.e., average fusion and adaptive fusion. For each fusion method, we test different settings including the fusions of two feature maps, three feature maps and four feature maps respectively against the case of only one single feature map for prediction. Results are listed in Table \ref{tab:scale}. As it can be seen, for averaging fusion, the case of three feature maps outperforms the others in terms of clipped Jaccard similarity and Dice coefficient, although the improvement is limited when compared to the one with a single feature map. One possible reason behind this is that the overall performance of the averaging feature map could be degraded when features of some specific scales have poor performance. Unlike average fusion, a great increase on performance can be seen for cases of adaptive fusion. For instance, the mean clipped Jaccard similarity increases from 73.12 to 77.07 when compared to the counterpart of average fusion. This demonstrates that fusion methods play a key role in applying multi-scale features for prediction.

\begin{table*}[]
\centering
\begin{threeparttable}
\caption{Effect of multi-scale/scale-adaptive features on the performance of the U-Net.}
\label{tab:scale}
\begin{tabular}{l|c|c|c|c|c|c}
\Xhline{3\arrayrulewidth}
\# scale & $SP$ & $PC$ & $RC$ & $DC$ & $JS$ & $S_{wsi}$ \\ \Xhline{2\arrayrulewidth}
1 (single) & 94.93 $\pm$ 1.45 & 89.29 $\pm$ 2.46 & 87.26 $\pm$ 5.76 & 88.20 $\pm$ 3.86 & 79.10 $\pm$ 6.12  & 72.96 $\pm$ 6.73 \\ \hline
2 (multiple) & \textbf{95.32 $\pm$ 1.41} & \textbf{89.92 $\pm$ 2.26} & 85.89 $\pm$ 7.23 & 87.70 $\pm$ 4.20 & 78.35 $\pm$ 6.63 & 73.12 $\pm$ 6.52\\
3 (multiple)  & 94.67 $\pm$ 1.68 & 88.96 $\pm$ 2.62 & \textbf{88.46 $\pm$ 6.46} & \textbf{88.61 $\pm$ 4.11} & \textbf{79.78 $\pm$ 6.46} & \textbf{75.40 $\pm$ 7.81}\\
4 (multiple) & 94.86 $\pm$ 1.36 & 89.10 $\pm$ 2.49 & 87.33 $\pm$ 5.80 & 88.16 $\pm$ 4.05 & 79.05 $\pm$ 6.38  & 73.26 $\pm$ 8.32\\ \hline
2 (adaptive) & 94.87 $\pm$ 2.00 & \textbf{89.54 $\pm$ 3.10} & \textbf{89.73 $\pm$ 6.46} & \textbf{89.52 $\pm$ 4.10} & \textbf{81.28 $\pm$ 6.56}  & 77.07 $\pm$ 8.62\\ 
3 (adaptive) & \textbf{94.96 $\pm$ 1.15} & 89.42 $\pm$ 2.50 & 89.22 $\pm$ 7.16 & 89.25 $\pm$ 4.81 & 80.91 $\pm$ 7.56  & \textbf{77.59 $\pm$ 8.92}\\ 
\Xhline{3\arrayrulewidth}
\end{tabular}
\end{threeparttable}
\end{table*}

\begin{figure*}
\centering
\includegraphics[width=1\textwidth]{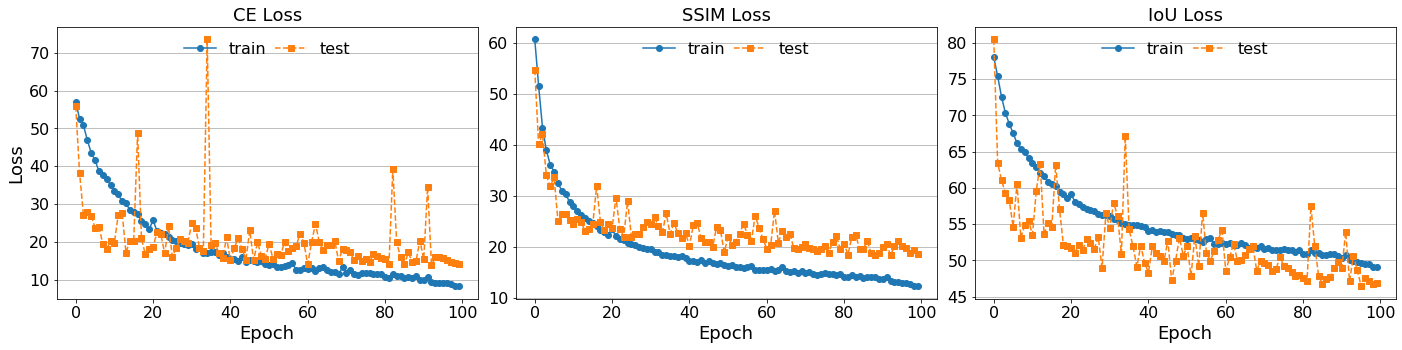}
\caption{\label{fig:losses} Losses against training epochs of the baseline U-Net.}
\end{figure*}

To demonstrate the influence of using different loss functions, We test networks trained by different losses including the widely used cross-entropy loss (CE), the SSIM loss, the IoU loss and different combinations. Table \ref{tab:loss} gives a comparison among these losses in terms of segmentation performance. An interesting observation is that either the SSIM loss or the IoU loss cannot work properly when applied alone. It is better to combine them with the CE loss. Improvements can be seen on the combination between CE and SSIM or combination of all three losses. Fig. \ref{fig:losses} shows the descent of each loss function in comparison between the training and testing modes. The objective function is to minimise the difference between a segmentation map and a corresponding ground truth. Since the CE loss is calculated on the pixel level, while SSIM is calculated on patch level and IoU is a global measurement, the final convergent value of each loss is different. 

\begin{table*}[]
\centering
\begin{threeparttable}
\caption{Effect of different loss functions on the performance of the baseline model U-Net.}
\label{tab:loss}
\begin{tabular}{l|c|c|c|c|c|c}
\Xhline{3\arrayrulewidth}
Loss function & $SP$ & $PC$ & $RC$ & $DC$ & $JS$ & $S_{wsi}$ \\ \Xhline{2\arrayrulewidth}
$\mathcal{L}_{ce}$ & 95.94 $\pm$ 1.08 & 91.22 $\pm$ 1.90 & 87.61 $\pm$ 7.02 & 89.24 $\pm$ 4.19 & 80.82 $\pm$ 6.68 & 73.78 $\pm$ 5.15\\
$\mathcal{L}_{iou}$ & 88.65 $\pm$ 3.99 & 79.22 $\pm$ 7.52 & \textbf{90.36 $\pm$ 4.89} & 84.36 $\pm$ 6.24 & 73.45 $\pm$ 9.24 & 56.89 $\pm$ 9.42\\
$\mathcal{L}_{ssim}$ & 95.56 $\pm$ 1.24 & 90.21 $\pm$ 2.58 & 85.23 $\pm$ 7.98 & 87.50 $\pm$ 5.11 & 78.14 $\pm$ 7.90 & 68.98 $\pm$ 4.97\\
$\mathcal{L}_{ce}$+$\mathcal{L}_{iou}$ & 95.58 $\pm$ 1.22 & 90.42 $\pm$ 2.70 & 88.00 $\pm$ 6.10 & 89.14 $\pm$ 4.32 & 80.67 $\pm$ 6.90  & 69.02 $\pm$ 8.77\\ 
$\mathcal{L}_{ce}$+$\mathcal{L}_{ssim}$ & \textbf{96.10 $\pm$ 0.74} & \textbf{91.48 $\pm$ 1.25} & 86.96 $\pm$ 6.66 & 89.04 $\pm$ 3.78 & 80.45 $\pm$ 6.07  & \textbf{75.63 $\pm$ 5.46}\\ 
$\mathcal{L}_{ce}$+$\mathcal{L}_{ssim}$+$\mathcal{L}_{iou}$ & 95.64 $\pm$ 0.49 & 90.67 $\pm$ 0.89 & 88.46 $\pm$ 6.64 & \textbf{89.42 $\pm$ 3.58} & \textbf{81.06 $\pm$ 5.78}  & 73.56 $\pm$ 2.97\\ 
\Xhline{3\arrayrulewidth}
\end{tabular}
\end{threeparttable}
\end{table*}

\subsection{Evaluation on liver cancer segmentation}
\label{sec:paip2019}
\begin{figure*}
\centering
\includegraphics[width=1\textwidth]{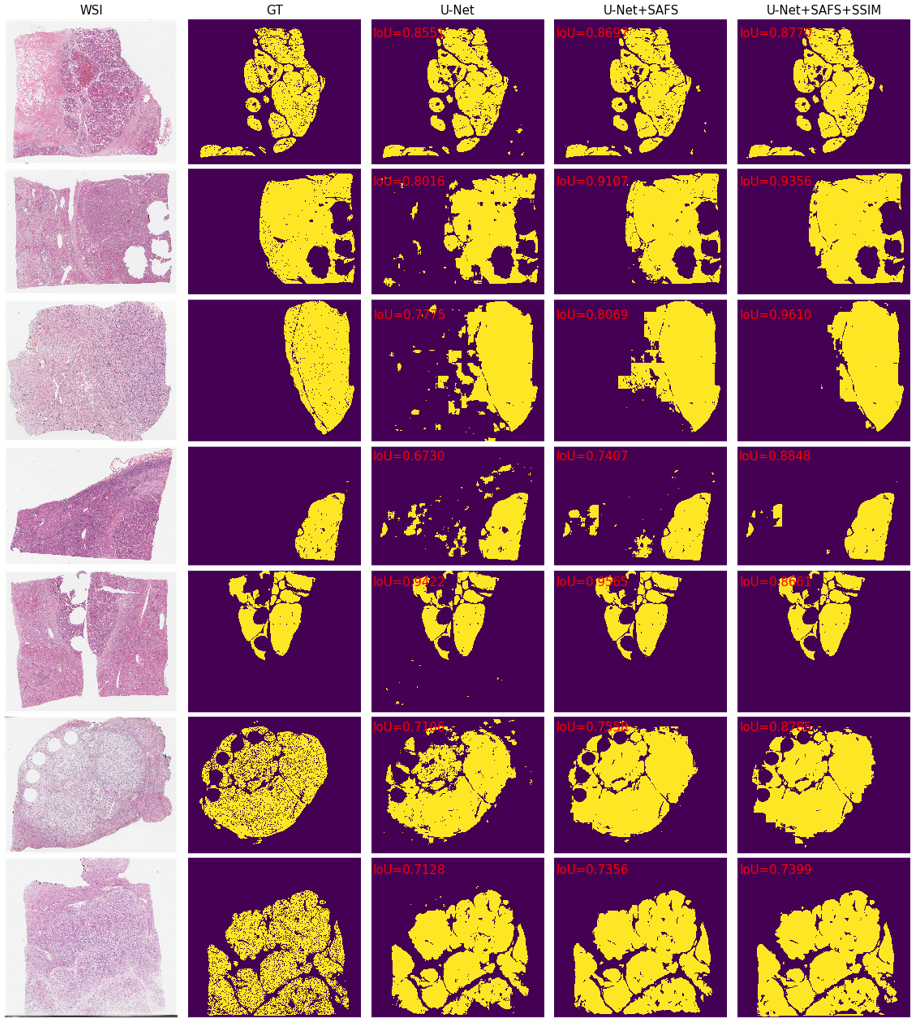}
\caption{\label{fig:segmap_liver} WSIs of liver cancer and segmentation maps of different networks.}
\end{figure*}

In this section, we test and compare networks with different architectures on liver cancer segmentation.
To validate the effectiveness of the proposed networks for cancer segmentation in whole slide images, digital slides of liver cancer from the challenge of PAIP 2019 are used for networks training and evaluation.
Besides the structure-aware scale-adaptive networks, several networks with superior selective-kernel convolutions and attention mechanisms including channel and spatial attention are also tested. Comparison among these networks are listed in Table \ref{tab:model}. The network (U-Net+SAFS+SSIM), to which both the SAFS module and the structural similarity (SSIM) are applied, scores a clipped Jaccard similarity of 79.89$\pm$5.08, with a 7\% improvement compared to the baseline U-Net.
The improvement can also be seen in Fig. \ref{fig:segmap_liver}, where many false positives of the baseline U-Net are gradually eliminated when the SAFS and the SSIM are applied, showing the effectiveness of the scale-adaptive features and structural similarity. The elimination of false positives, most of which are dissociated from viable tumour areas, is more obvious while the SSIM is applied, due to the ability of SSIM to focus on differences of local patches.

The scale-adaptive feature selection module and its combinations with the structural similarity (SSIM), shortcut convolution (SC), and selective-kernel convolution (SK) are tested respectively, results listed in Table \ref{tab:team} together with the top ten results of the PAIP 2019 challenge. Note that a testing set of 40 extra WSIs are provided during the period of the PAIP 2019, while our results are cross-validated using the training set. Although it is not absolutely fair to make such a comparison and the training set we use is smaller, the results are listed for reference. As it can be seen, the proposed networks achieve outstanding performance. Specifically, the network with the combination of SAFS and SSIM (U-Net+SAFS+SSIM) scores $ 79.89\pm 5.08$ which is almost the same as the best published result of an ensemble model\cite{wang2021hybrid}, while the network with shortcut and selective-kernel convolutions (U-Net+SK+SC) achieves the best clipped Jaccard similarity of 81.00$\pm$4.98.
Although the proposed network achieves a high overall performance, the variance is merely acceptable. The relatively large variance reveals that our network may face generalisation issues to some instances. We will pay special attention to these issues in our future work.

\begin{table*}[]
\centering
\setlength\tabcolsep{5pt}
\begin{threeparttable}
\caption{Comparison of networks with different architectures tested on the PAIP 2019 and the PAIP 2020 respectively. \#P denotes the number of parameters.}
\label{tab:model}
\begin{tabular}{l|r|ccc|ccc}
\Xhline{3\arrayrulewidth}
\multirow{2}{*}{Architecture} & \multicolumn{1}{c|}{\multirow{2}{*}{\begin{tabular}[c]{@{}c@{}}\#P\\ (million)\end{tabular}}} & \multicolumn{3}{c|}{PAIP 2019 (liver cancer)} & \multicolumn{3}{c}{PAIP 2020 (colorectal cancer)} \\ \cline{3-8} 
 & \multicolumn{1}{c|}{} & $DC$ & $JS$ & $S_{wsi}$ & $DC$ & $JS$ & $S_{wsi}$ \\ \Xhline{2\arrayrulewidth}
U-Net (baseline) & \textbf{8.7680} & 88.20 $\pm$ 3.86 & 79.10 $\pm$ 6.12 & 72.96 $\pm$ 6.73 & 91.72 $\pm$ 3.15 & 84.86 $\pm$ 5.30 & 80.12 $\pm$ 7.87 \\ 
U-Net+SC & 8.9713 & 88.31 $\pm$ 4.28 & 79.33 $\pm$ 6.75 & 73.05 $\pm$ 6.45 & 91.87 $\pm$ 3.19 & 85.12 $\pm$ 5.39 & 79.53 $\pm$ 6.47 \\ 
U-Net+AG & 8.8826 & 88.80 $\pm$ 4.45 & 80.14 $\pm$ 7.06 & 75.77 $\pm$ 8.10 & 92.32 $\pm$ 3.09 & 85.88 $\pm$ 5.25 & 79.84 $\pm$ 8.76 \\ 
U-Net+SE & 8.9922 & 89.48 $\pm$ 4.62 & 81.27 $\pm$ 7.29 & 79.02 $\pm$ 5.04 & 92.64 $\pm$ 2.99 & 86.44 $\pm$ 5.12 & 83.10 $\pm$ 8.16 \\ 
U-Net+BAM & 9.3177 & 90.00 $\pm$ 3.72 & 82.01 $\pm$ 6.04 & 76.14 $\pm$ 8.95 & \textbf{93.01 $\pm$ 2.85} & \textbf{87.06 $\pm$ 4.95} & 83.22 $\pm$ 9.42 \\ 
U-Net+CBAM & 8.9963 & 87.75 $\pm$ 3.90 & 78.39 $\pm$ 6.25 & 75.08 $\pm$ 5.53 & 92.78 $\pm$ 2.60 & 86.64 $\pm$ 4.52 & 83.00 $\pm$ 8.28 \\ 
U-Net+SK & 19.8725 & 89.51 $\pm$ 4.48 & 81.30 $\pm$ 7.17 & 79.11 $\pm$ 6.98 & 92.84 $\pm$ 2.58 & 86.74 $\pm$ 4.49 & 82.06 $\pm$ 8.65 \\ 
U-Net+MS & 8.8791 & 88.61 $\pm$ 4.11 & 79.78 $\pm$ 6.46 & 75.40 $\pm$ 7.81 & 91.87 $\pm$ 3.16 & 85.12 $\pm$ 5.36 & 80.05 $\pm$ 7.94 \\
U-Net+SAFS & 8.8051 & 89.52 $\pm$ 4.10 & 81.28 $\pm$ 6.56  & 77.07 $\pm$ 8.62 & 92.42 $\pm$ 3.21 & 86.08 $\pm$ 5.50 & 82.97 $\pm$ 4.97 \\
U-Net+SAFS+SSIM & 8.8051 & \textbf{90.06 $\pm$ 3.65} & \textbf{82.11 $\pm$ 5.93}  & \textbf{79.89 $\pm$ 5.08} & 92.46 $\pm$ 3.18 & 86.14 $\pm$ 5.43 & \textbf{84.22 $\pm$ 5.22} \\
\Xhline{3\arrayrulewidth}
\end{tabular}
\end{threeparttable}
\end{table*}

\begin{table}[]
\centering
\begin{threeparttable}
\caption{Comparison with top ten teams participating in the challenge of PAIP 2019. Ranks are from the challenge leaderboard \cite{kim2021paip}.}
\label{tab:team}
\begin{tabular}{lll}
\Xhline{3\arrayrulewidth}
Rank & Team & $S_{wsi}$ \\
\Xhline{2\arrayrulewidth}
- & U-Net+SAFS & 77.07$\pm$8.62 \\
- & U-Net+SAFS+SC & 78.66$\pm$5.85 \\
- & U-Net+SAFS+SSIM & 79.89$\pm$5.08 \\
- & U-Net+SAFS+SK & 79.99$\pm$8.01 \\
- & U-Net+SK+SC & \textbf{81.00$\pm$4.98} \\
- & Ensemble \cite{wang2021hybrid} & 79.70 \\
1 & Hyun Jung & 78.90 \\
2 & Team Sen & 77.72 \\
3 & Team MIRL-IITM & 75.03 \\
4 & Team Damo AIC & 67.18 \\
5 & Team QuIIL & 66.52 \\
6 & Team CUHK-Med & 66.24 \\
7 & Team DAISYlab@UKE & 65.96 \\
8 & Team COSYPath & 63.13 \\
9 & Ching-Wei Wang & 60.65 \\
10 & Team LRDE & 52.99 \\
\Xhline{3\arrayrulewidth}
\end{tabular}
\end{threeparttable}
\end{table}

\subsection{Evaluation on colorectal cancer segmentation}
\begin{figure*}
\centering
\includegraphics[width=1\textwidth]{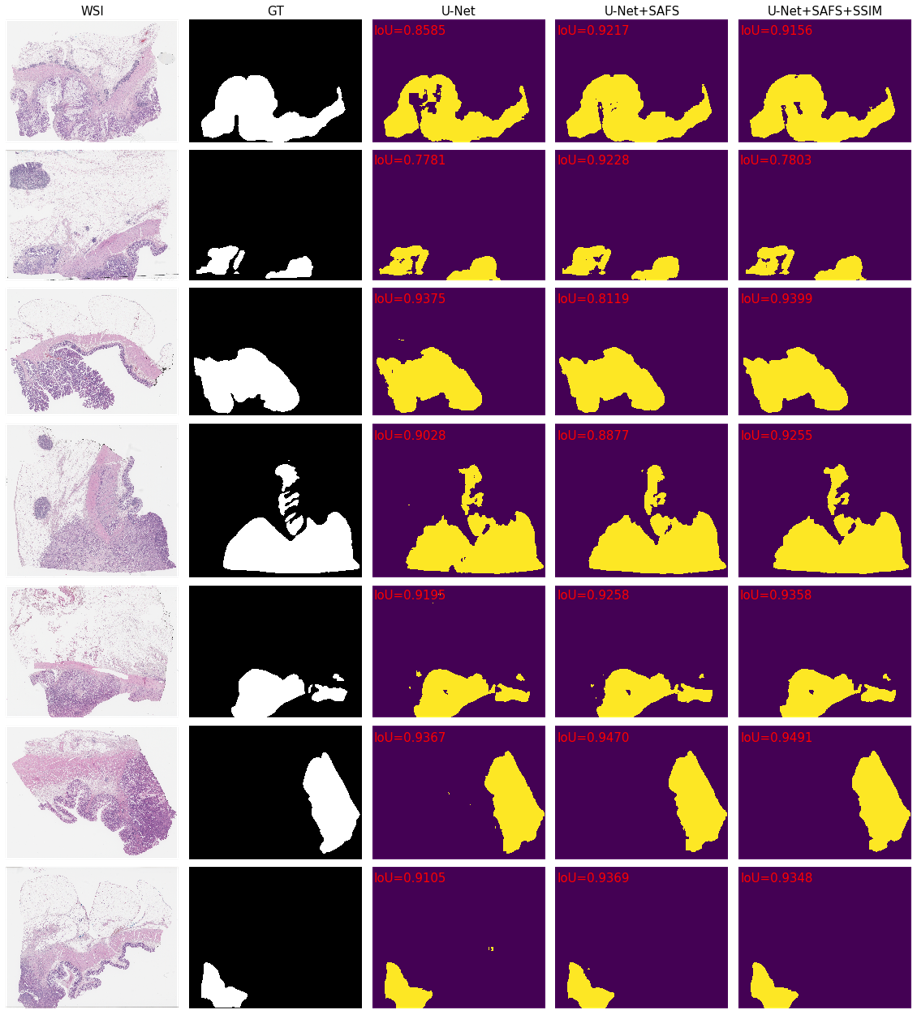}
\caption{\label{fig:segmap_colo} WSIs of colorectal cancer and segmentation maps of different networks.}
\end{figure*}

In this section, we further test and compare the proposed networks on colorectal cancer segmentation using WSIs of PAIP 2020 in which tiny blank regions within viable tumour areas are also considered as cancer (see the GT column in Fig. \ref{fig:segmap_colo}). Results are listed in Table \ref{tab:model}. An interesting observation is that either the scale-adaptive feature selection module or the structural similarity can lead to improvement in colorectal cancer segmentation when compared to the baseline, which is similar to the evaluation on liver cancer segmentation. Especially when both methods are applied simultaneously, the network (U-Net+SAFS+SSIM) achieves the best clipped Jaccard similarity of 84.22$\pm$5.22.
We visualize some examples of whole-slide images and the corresponding segmentation maps from networks with different settings to demonstrate the improvement (Fig. \ref{fig:segmap_colo}).
However, the improvement is not as significant as the results on the liver cancer task, mainly because there is less influences of small regions outside the viable tumour areas in WSIs of colorectal cancer.
Networks with superior attention mechanisms can also outperform the baseline. For instance, U-Net+BAM scores better in Jaccard similarity (JS) and Dice coefficient (DC) on small image patches. Nevertheless, considering the tradeoff between the number of network parameters and segmentation performance, we propose the combination of SAFS and SSIM for efficient and accurate cancer segmentation.

\section{Conclusion}
We present a structure-aware scale-adaptive network to efficiently and accurately segment tumour area in whole-slide images. This challenging image segmentation problem is tackled by applying a SAFS module to dynamically select feature maps from different scales for final prediction. Moreover, the structural similarity metric is proposed to regulate the network training procedure for better tissue structure capturing, so as to improve the segmentation performance without increasing the number of network parameters. The proposed network has been evaluated on two cancer segmentation benchmarks. Significant improvements are achieved by exploiting scale-adaptive features together with the structural similarity loss, demonstrating the effectiveness of the dynamic and scale-adaptive feature selection approach for cancer segmentation. Specifically, experimental results show that the proposed network yield outstanding performance on the benchmark of PAIP 2019 for liver cancer segmentation while compared to previous methods, and on the colorectal cancer segmentation benchmark when compared with the baseline or other networks with excellent attention designs.
Future work will further investigate and compare the proposed networks for cancer segmentation on WSIs at different magnification scales.

\bibliographystyle{IEEEbib}
\bibliography{seg}

\begin{thebibliography}{10}

\bibitem{sung2021global}
Hyuna Sung, Jacques Ferlay, Rebecca~L Siegel, Mathieu Laversanne, Isabelle
  Soerjomataram, Ahmedin Jemal, and Freddie Bray,
\newblock ``Global cancer statistics 2020: Globocan estimates of incidence and
  mortality worldwide for 36 cancers in 185 countries,''
\newblock {\em CA: a cancer journal for clinicians}, vol. 71, no. 3, pp.
  209--249, 2021.

\bibitem{tolkachev2020deep}
Alexey Tolkachev, Ilyas Sirazitdinov, Maksym Kholiavchenko, Tamerlan Mustafaev,
  and Bulat Ibragimov,
\newblock ``Deep learning for diagnosis and segmentation of pneumothorax: the
  results on the kaggle competition and validation against radiologists,''
\newblock {\em IEEE Journal of Biomedical and Health Informatics}, vol. 25, no.
  5, pp. 1660--1672, 2020.

\bibitem{shen2017fully}
Haocheng Shen and Jianguo Zhang,
\newblock ``Fully connected crf with data-driven prior for multi-class brain
  tumor segmentation,''
\newblock in {\em 2017 IEEE International Conference on Image Processing
  (ICIP)}. IEEE, 2017, pp. 1727--1731.

\bibitem{akil2020fully}
Mohamed Akil, Rachida Saouli, Rostom Kachouri, et~al.,
\newblock ``Fully automatic brain tumor segmentation with deep learning-based
  selective attention using overlapping patches and multi-class weighted
  cross-entropy,''
\newblock {\em Medical image analysis}, vol. 63, pp. 101692, 2020.

\bibitem{lei2020skin}
Baiying Lei, Zaimin Xia, Feng Jiang, Xudong Jiang, Zongyuan Ge, Yanwu Xu, Jing
  Qin, Siping Chen, Tianfu Wang, and Shuqiang Wang,
\newblock ``Skin lesion segmentation via generative adversarial networks with
  dual discriminators,''
\newblock {\em Medical Image Analysis}, vol. 64, pp. 101716, 2020.

\bibitem{alba2018automatic}
Xenia Alba, Karim Lekadir, Marco Pereanez, Pau Medrano-Gracia, Alistair~A
  Young, and Alejandro~F Frangi,
\newblock ``Automatic initialization and quality control of large-scale cardiac
  mri segmentations,''
\newblock {\em Medical image analysis}, vol. 43, pp. 129--141, 2018.

\bibitem{schlemper2019attention}
Jo~Schlemper, Ozan Oktay, Michiel Schaap, Mattias Heinrich, Bernhard Kainz, Ben
  Glocker, and Daniel Rueckert,
\newblock ``Attention gated networks: Learning to leverage salient regions in
  medical images,''
\newblock {\em Medical image analysis}, vol. 53, pp. 197--207, 2019.

\bibitem{schmitz2021multi}
R{\"u}diger Schmitz, Frederic Madesta, Maximilian Nielsen, Jenny Krause, Stefan
  Steurer, Ren{\'e} Werner, and Thomas R{\"o}sch,
\newblock ``Multi-scale fully convolutional neural networks for histopathology
  image segmentation: from nuclear aberrations to the global tissue
  architecture,''
\newblock {\em Medical image analysis}, vol. 70, pp. 101996, 2021.

\bibitem{ronneberger2015u}
Olaf Ronneberger, Philipp Fischer, and Thomas Brox,
\newblock ``U-net: Convolutional networks for biomedical image segmentation,''
\newblock in {\em International Conference on Medical image computing and
  computer-assisted intervention}. Springer, 2015, pp. 234--241.

\bibitem{he2016deep}
Kaiming He, Xiangyu Zhang, Shaoqing Ren, and Jian Sun,
\newblock ``Deep residual learning for image recognition,''
\newblock in {\em Proceedings of the IEEE conference on computer vision and
  pattern recognition}, 2016, pp. 770--778.

\bibitem{farabet2012learning}
Clement Farabet, Camille Couprie, Laurent Najman, and Yann LeCun,
\newblock ``Learning hierarchical features for scene labeling,''
\newblock {\em IEEE transactions on pattern analysis and machine intelligence},
  vol. 35, no. 8, pp. 1915--1929, 2012.

\bibitem{gatta2014unrolling}
Carlo Gatta, Adriana Romero, and Joost van~de Veijer,
\newblock ``Unrolling loopy top-down semantic feedback in convolutional deep
  networks,''
\newblock in {\em Proceedings of the IEEE Conference on Computer Vision and
  Pattern Recognition Workshops}, 2014, pp. 498--505.

\bibitem{grangier2009deep}
David Grangier, L{\'e}on Bottou, and Ronan Collobert,
\newblock ``Deep convolutional networks for scene parsing,''
\newblock in {\em ICML 2009 Deep Learning Workshop}. Citeseer, 2009, vol.~3, p.
  109.

\bibitem{long2015fully}
Jonathan Long, Evan Shelhamer, and Trevor Darrell,
\newblock ``Fully convolutional networks for semantic segmentation,''
\newblock in {\em Proceedings of the IEEE conference on computer vision and
  pattern recognition}, 2015, pp. 3431--3440.

\bibitem{badrinarayanan2017segnet}
Vijay Badrinarayanan, Alex Kendall, and Roberto Cipolla,
\newblock ``Segnet: A deep convolutional encoder-decoder architecture for image
  segmentation,''
\newblock {\em IEEE transactions on pattern analysis and machine intelligence},
  vol. 39, no. 12, pp. 2481--2495, 2017.

\bibitem{noh2015learning}
Hyeonwoo Noh, Seunghoon Hong, and Bohyung Han,
\newblock ``Learning deconvolution network for semantic segmentation,''
\newblock in {\em Proceedings of the IEEE international conference on computer
  vision}, 2015, pp. 1520--1528.

\bibitem{chen2017deeplab}
Liang-Chieh Chen, George Papandreou, Iasonas Kokkinos, Kevin Murphy, and Alan~L
  Yuille,
\newblock ``Deeplab: Semantic image segmentation with deep convolutional nets,
  atrous convolution, and fully connected crfs,''
\newblock {\em IEEE transactions on pattern analysis and machine intelligence},
  vol. 40, no. 4, pp. 834--848, 2017.

\bibitem{zhou2018unet++}
Zongwei Zhou, Md~Mahfuzur~Rahman Siddiquee, Nima Tajbakhsh, and Jianming Liang,
\newblock ``Unet++: A nested u-net architecture for medical image
  segmentation,''
\newblock in {\em Deep learning in medical image analysis and multimodal
  learning for clinical decision support}, pp. 3--11. Springer, 2018.

\bibitem{li2019cascade}
Suiyi Li, Yuxuan Chen, Su~Yang, and Wuyang Luo,
\newblock ``Cascade dense-unet for prostate segmentation in mr images,''
\newblock in {\em International Conference on Intelligent Computing}. Springer,
  2019, pp. 481--490.

\bibitem{li2020attention}
Chen Li, Yusong Tan, Wei Chen, Xin Luo, Yuanming Gao, Xiaogang Jia, and Zhiying
  Wang,
\newblock ``Attention unet++: A nested attention-aware u-net for liver ct image
  segmentation,''
\newblock in {\em 2020 IEEE International Conference on Image Processing
  (ICIP)}. IEEE, 2020, pp. 345--349.

\bibitem{li2020net}
Yilong Li, Zhaoyang Xu, Yaqi Wang, Huiyu Zhou, and Qianni Zhang,
\newblock ``Su-net and du-net fusion for tumour segmentation in histopathology
  images,''
\newblock in {\em 2020 IEEE 17th International Symposium on Biomedical Imaging
  (ISBI)}. IEEE, 2020, pp. 461--465.

\bibitem{feng2020deep}
Ruiwei Feng, Xuechen Liu, Jintai Chen, Danny~Z Chen, Honghao Gao, and Jian Wu,
\newblock ``A deep learning approach for colonoscopy pathology wsi analysis:
  accurate segmentation and classification,''
\newblock {\em IEEE Journal of Biomedical and Health Informatics}, 2020.

\bibitem{shah2021colorectal}
Nisarg~A Shah, Divij Gupta, Romil Lodaya, Ujjwal Baid, and Sanjay Talbar,
\newblock ``Colorectal cancer segmentation using atrous convolution and
  residual enhanced unet,''
\newblock {\em arXiv preprint arXiv:2103.09289}, 2021.

\bibitem{janowczyk2016deep}
Andrew Janowczyk and Anant Madabhushi,
\newblock ``Deep learning for digital pathology image analysis: A comprehensive
  tutorial with selected use cases,''
\newblock {\em Journal of pathology informatics}, vol. 7, 2016.

\bibitem{litjens2017survey}
Geert Litjens, Thijs Kooi, Babak~Ehteshami Bejnordi, Arnaud Arindra~Adiyoso
  Setio, Francesco Ciompi, Mohsen Ghafoorian, Jeroen~Awm Van Der~Laak, Bram
  Van~Ginneken, and Clara~I S{\'a}nchez,
\newblock ``A survey on deep learning in medical image analysis,''
\newblock {\em Medical image analysis}, vol. 42, pp. 60--88, 2017.

\bibitem{araujo2017classification}
Teresa Ara{\'u}jo, Guilherme Aresta, Eduardo Castro, Jos{\'e} Rouco, Paulo
  Aguiar, Catarina Eloy, Ant{\'o}nio Pol{\'o}nia, and Aur{\'e}lio Campilho,
\newblock ``Classification of breast cancer histology images using
  convolutional neural networks,''
\newblock {\em PloS one}, vol. 12, no. 6, pp. e0177544, 2017.

\bibitem{sharma2017deep}
Harshita Sharma, Norman Zerbe, Iris Klempert, Olaf Hellwich, and Peter Hufnagl,
\newblock ``Deep convolutional neural networks for automatic classification of
  gastric carcinoma using whole slide images in digital histopathology,''
\newblock {\em Computerized Medical Imaging and Graphics}, vol. 61, pp. 2--13,
  2017.

\bibitem{ciresan2012deep}
Dan Ciresan, Alessandro Giusti, Luca Gambardella, and J{\"u}rgen Schmidhuber,
\newblock ``Deep neural networks segment neuronal membranes in electron
  microscopy images,''
\newblock {\em Advances in neural information processing systems}, vol. 25, pp.
  2843--2851, 2012.

\bibitem{jahromi2020deep}
Seyed Alireza~Fatemi Jahromi, Ali~Asghar Khani, Hatef~Otroshi Shahreza,
  Mahdieh~Soleymani Baghshah, and Hamid Behroozi,
\newblock ``A deep learning framework for viable tumor burden estimation,''
\newblock in {\em 2020 6th Iranian Conference on Signal Processing and
  Intelligent Systems (ICSPIS)}. IEEE, 2020, pp. 1--7.

\bibitem{wang2021hybrid}
Xiyue Wang, Yuqi Fang, Sen Yang, Delong Zhu, Minghui Wang, Jing Zhang, Kai-yu
  Tong, and Xiao Han,
\newblock ``A hybrid network for automatic hepatocellular carcinoma
  segmentation in h\&e-stained whole slide images,''
\newblock {\em Medical Image Analysis}, vol. 68, pp. 101914, 2021.

\bibitem{itti1998model}
Laurent Itti, Christof Koch, and Ernst Niebur,
\newblock ``A model of saliency-based visual attention for rapid scene
  analysis,''
\newblock {\em IEEE Transactions on pattern analysis and machine intelligence},
  vol. 20, no. 11, pp. 1254--1259, 1998.

\bibitem{rensink2000dynamic}
Ronald~A Rensink,
\newblock ``The dynamic representation of scenes,''
\newblock {\em Visual cognition}, vol. 7, no. 1-3, pp. 17--42, 2000.

\bibitem{lecun2015deep}
Yann LeCun, Yoshua Bengio, and Geoffrey Hinton,
\newblock ``Deep learning,''
\newblock {\em nature}, vol. 521, no. 7553, pp. 436--444, 2015.

\bibitem{krizhevsky2012imagenet}
Alex Krizhevsky, Ilya Sutskever, and Geoffrey~E Hinton,
\newblock ``Imagenet classification with deep convolutional neural networks,''
\newblock {\em Advances in neural information processing systems}, vol. 25, pp.
  1097--1105, 2012.

\bibitem{szegedy2015going}
Christian Szegedy, Wei Liu, Yangqing Jia, Pierre Sermanet, Scott Reed, Dragomir
  Anguelov, Dumitru Erhan, Vincent Vanhoucke, and Andrew Rabinovich,
\newblock ``Going deeper with convolutions,''
\newblock in {\em Proceedings of the IEEE conference on computer vision and
  pattern recognition}, 2015, pp. 1--9.

\bibitem{fu2017look}
Jianlong Fu, Heliang Zheng, and Tao Mei,
\newblock ``Look closer to see better: Recurrent attention convolutional neural
  network for fine-grained image recognition,''
\newblock in {\em Proceedings of the IEEE conference on computer vision and
  pattern recognition}, 2017, pp. 4438--4446.

\bibitem{hu2018squeeze}
Jie Hu, Li~Shen, and Gang Sun,
\newblock ``Squeeze-and-excitation networks,''
\newblock in {\em Proceedings of the IEEE conference on computer vision and
  pattern recognition}, 2018, pp. 7132--7141.

\bibitem{fu2019dual}
Jun Fu, Jing Liu, Haijie Tian, Yong Li, Yongjun Bao, Zhiwei Fang, and Hanqing
  Lu,
\newblock ``Dual attention network for scene segmentation,''
\newblock in {\em Proceedings of the IEEE/CVF Conference on Computer Vision and
  Pattern Recognition}, 2019, pp. 3146--3154.

\bibitem{mi2020hierarchical}
Li~Mi and Zhenzhong Chen,
\newblock ``Hierarchical graph attention network for visual relationship
  detection,''
\newblock in {\em Proceedings of the IEEE/CVF Conference on Computer Vision and
  Pattern Recognition}, 2020, pp. 13886--13895.

\bibitem{guo2021ssan}
Xudong Guo, Xun Guo, and Yan Lu,
\newblock ``Ssan: Separable self-attention network for video representation
  learning,''
\newblock in {\em Proceedings of the IEEE/CVF Conference on Computer Vision and
  Pattern Recognition}, 2021, pp. 12618--12627.

\bibitem{park2018bam}
Jongchan Park, Sanghyun Woo, Joon-Young Lee, and In~So Kweon,
\newblock ``Bam: Bottleneck attention module,''
\newblock {\em arXiv preprint arXiv:1807.06514}, 2018.

\bibitem{woo2018cbam}
Sanghyun Woo, Jongchan Park, Joon-Young Lee, and In~So Kweon,
\newblock ``Cbam: Convolutional block attention module,''
\newblock in {\em Proceedings of the European conference on computer vision
  (ECCV)}, 2018, pp. 3--19.

\bibitem{oktay2018attention}
Ozan Oktay, Jo~Schlemper, Loic~Le Folgoc, Matthew Lee, Mattias Heinrich,
  Kazunari Misawa, Kensaku Mori, Steven McDonagh, Nils~Y Hammerla, Bernhard
  Kainz, et~al.,
\newblock ``Attention u-net: Learning where to look for the pancreas,''
\newblock {\em arXiv preprint arXiv:1804.03999}, 2018.

\bibitem{ioffe2015batch}
Sergey Ioffe and Christian Szegedy,
\newblock ``Batch normalization: Accelerating deep network training by reducing
  internal covariate shift,''
\newblock in {\em International conference on machine learning}. PMLR, 2015,
  pp. 448--456.

\bibitem{nair2010rectified}
Vinod Nair and Geoffrey~E Hinton,
\newblock ``Rectified linear units improve restricted boltzmann machines,''
\newblock in {\em Icml}, 2010.

\bibitem{kim2021paip}
Yoo~Jung Kim, Hyungjoon Jang, Kyoungbun Lee, Seongkeun Park, Sung-Gyu Min,
  Choyeon Hong, Jeong~Hwan Park, Kanggeun Lee, Jisoo Kim, Wonjae Hong, et~al.,
\newblock ``Paip 2019: Liver cancer segmentation challenge,''
\newblock {\em Medical Image Analysis}, vol. 67, pp. 101854, 2021.

\bibitem{li2019selective}
Xiang Li, Wenhai Wang, Xiaolin Hu, and Jian Yang,
\newblock ``Selective kernel networks,''
\newblock in {\em Proceedings of the IEEE/CVF Conference on Computer Vision and
  Pattern Recognition}, 2019, pp. 510--519.

\bibitem{lin2017feature}
Tsung-Yi Lin, Piotr Doll{\'a}r, Ross Girshick, Kaiming He, Bharath Hariharan,
  and Serge Belongie,
\newblock ``Feature pyramid networks for object detection,''
\newblock in {\em Proceedings of the IEEE conference on computer vision and
  pattern recognition}, 2017, pp. 2117--2125.

\bibitem{nguyen2019frequency}
Thanh Nguyen-Duc, Tran~Minh Quan, and Won-Ki Jeong,
\newblock ``Frequency-splitting dynamic mri reconstruction using multi-scale 3d
  convolutional sparse coding and automatic parameter selection,''
\newblock {\em Medical image analysis}, vol. 53, pp. 179--196, 2019.

\bibitem{zhou2004image}
Wang Zhou,
\newblock ``Image quality assessment: from error measurement to structural
  similarity,''
\newblock {\em IEEE transactions on image processing}, vol. 13, pp. 600--613,
  2004.

\bibitem{wang2003multiscale}
Zhou Wang, Eero~P Simoncelli, and Alan~C Bovik,
\newblock ``Multiscale structural similarity for image quality assessment,''
\newblock in {\em The Thrity-Seventh Asilomar Conference on Signals, Systems \&
  Computers, 2003}. Ieee, 2003, vol.~2, pp. 1398--1402.

\bibitem{rahman2016optimizing}
Md~Atiqur Rahman and Yang Wang,
\newblock ``Optimizing intersection-over-union in deep neural networks for
  image segmentation,''
\newblock in {\em International symposium on visual computing}. Springer, 2016,
  pp. 234--244.

\bibitem{mattyus2017deeproadmapper}
Gell{\'e}rt M{\'a}ttyus, Wenjie Luo, and Raquel Urtasun,
\newblock ``Deeproadmapper: Extracting road topology from aerial images,''
\newblock in {\em Proceedings of the IEEE International Conference on Computer
  Vision}, 2017, pp. 3438--3446.

\bibitem{sudre2017generalised}
Carole~H Sudre, Wenqi Li, Tom Vercauteren, Sebastien Ourselin, and M~Jorge
  Cardoso,
\newblock ``Generalised dice overlap as a deep learning loss function for
  highly unbalanced segmentations,''
\newblock in {\em Deep learning in medical image analysis and multimodal
  learning for clinical decision support}, pp. 240--248. Springer, 2017.

\bibitem{kingma2014adam}
Diederik~P Kingma and Jimmy Ba,
\newblock ``Adam: A method for stochastic optimization,''
\newblock {\em arXiv preprint arXiv:1412.6980}, 2014.

\bibitem{tan2019efficientnet}
Mingxing Tan and Quoc Le,
\newblock ``Efficientnet: Rethinking model scaling for convolutional neural
  networks,''
\newblock in {\em International Conference on Machine Learning}. PMLR, 2019,
  pp. 6105--6114.

\end{thebibliography}

\end{document}